\documentclass[letterpaper, 10 pt, conference]{ieeeconf}  

\usepackage[top=54pt, bottom=54pt, left=0.75in, right=0.75in]{geometry}

\IEEEoverridecommandlockouts                              

\usepackage{tabularx}
\usepackage{array}

\usepackage{float} 

\usepackage{amssymb}
\usepackage{newtxtext,newtxmath} 
\usepackage{graphicx}
\usepackage[export]{adjustbox}
\usepackage{url}
\usepackage{latexsym}
\usepackage{algpseudocode}
\usepackage{amsmath}
\usepackage{lipsum}
\usepackage{caption}
\usepackage{bbding}
\usepackage{wasysym}
\usepackage{blindtext}
\usepackage{booktabs}
\usepackage{tabularx}
\usepackage{array}
\usepackage{pifont}
\usepackage[export]{adjustbox}
\usepackage{float}
\usepackage{multirow}

\usepackage{amsthm}
\usepackage{xcolor}
\usepackage{cite}

\title{\LARGE \bf
Towards Transparent Ethical AI: A Roadmap for 
Trustworthy Robotic Systems
}

\author{Ahmad Farooq*$^{1}$ and Kamran Iqbal$^{2}$
\thanks{This work was not supported by any organization.}
\thanks{*Corresponding Author: Ahmad Farooq}
\thanks{$^{1}$Ahmad Farooq is a Ph.D. Candidiate in the Electrical and Computer Engineering Department at the University of Arkansas at Little Rock, AR, 72204, USA
        {\tt\small afarooq@ualr.edu}}%
\thanks{$^{2}$Kamran Iqbal is a Professor in the Electrical and Computer Engineering Department at the University of Arkansas at Little Rock, AR, 72204, USA
        {\tt\small kxiqbal@ualr.edu}}%
\thanks{This is the author's preprint version of an article published in RCVE'25: Proceedings of the 2025 3rd International Conference on Robotics, Control and Vision Engineering. The final published version is available in the ACM Digital Library: https://doi.org/10.1145/3747393.3747399.}%
}

\begin{document}

\maketitle
\thispagestyle{empty}
\pagestyle{empty}

\begin{abstract}
As artificial intelligence (AI) and robotics increasingly permeate society, ensuring the ethical behavior of these systems has become paramount. This paper contends that transparency in AI decision-making processes is fundamental to developing trustworthy and ethically aligned robotic systems. We explore how transparency facilitates accountability, enables informed consent, and supports the debugging of ethical algorithms. The paper outlines technical, ethical, and practical challenges in implementing transparency and proposes novel approaches to enhance it, including standardized metrics, explainable AI techniques, and user-friendly interfaces. This paper introduces a framework that connects technical implementation with ethical considerations in robotic systems, focusing on the specific challenges of achieving transparency in dynamic, real-world contexts. We analyze how prioritizing transparency can impact public trust, regulatory policies, and avenues for future research. By positioning transparency as a fundamental element in ethical AI system design, we aim to add to the ongoing discussion on responsible AI and robotics, providing direction for future advancements in this vital field.

\end{abstract}

\section{Introduction}
The swift progress in artificial intelligence (AI) and robotics has brought about an era of autonomous systems with complex decision-making capabilities. The rapid pace of AI advancements, including large language models and autonomous systems, has highlighted the need for transparent and ethical AI decision-making. These robotic systems are now part of many aspects of daily life—ranging from healthcare and transportation to manufacturing and security—thereby raising important ethical questions about their decision-making processes \cite{winfield2018ethical, arrieta2020explainable}.

Achieving transparency in AI involves more than just making the code accessible. It requires comprehending, interpreting, and explaining the logic behind a system’s decisions and behaviors. In ethical robotics, transparency is not just a technical consideration but a fundamental ethical principle underpinning trust, accountability, and responsible innovation \cite{floridi2018ai4people, rudin2019stop}. However, current approaches to ethical AI in robotics often lack sufficient transparency, creating an opaque ``black box" that hinders our ability to verify ethical compliance and address potential biases or errors.

This opacity in AI decision-making processes poses significant challenges. It impedes our ability to ensure accountability, obtain informed consent from users and stakeholders, and effectively debug and improve ethical algorithms \cite{doshi2017towards, cave2019hopes}. Moreover, the lack of transparency can erode public trust in robotic systems, potentially slowing their adoption and limiting their societal benefits.

This paper argues that transparency should be elevated to a fundamental principle in the development of ethical robotic systems. By illuminating the decision-making processes of AI-driven robots, we can build trust, facilitate meaningful human oversight, and ultimately create more ethically aligned robotic systems \cite{bryson2017standardizing}. While acknowledging the technical, ethical, and practical hurdles in implementing transparency, this paper proposes innovative approaches to enhance it and discusses the far-reaching implications of this focus for the future of ethical AI in robotics.

The remainder of this paper is structured as follows: Section II presents the case for transparency in AI systems. Section III discusses the challenges in implementing transparency, while Section IV proposes approaches to enhance transparency in ethical AI decision-making. Section V examines the implications of prioritizing transparency and outlines future research directions. Finally, Section VI concludes the paper along with a call to action for the robotics and AI community.

\section{The Case for Transparency} 
Lipton \cite{lipton2018mythos} defines transparency as comprising both model interpretability and post-hoc explanations. In this paper, we adopt a broad, layered perspective of transparency, ensuring it is both technically accessible and comprehensible to users. Transparency in robotic systems refers to offering clear and comprehensible insights into how the system makes decisions—covering all aspects from the data inputs to the algorithms applied and the reasoning behind the outcomes. This idea transcends basic technical openness, requiring accessibility and interpretability for a wide range of stakeholders, including users, developers, policymakers, and ethicists \cite{wachter2017transparent}.

To determine if a system truly embodies transparency, we propose several criteria:

\begin{enumerate} \item \textbf{Algorithmic Transparency:} The ability to inspect and understand the core algorithms and data processing techniques that are applied within the system. 
\item \textbf{Functional Transparency:} Clear explanations of the system’s functions, limitations, and the intended use cases. 
\item \textbf{Operational Transparency:} Real-time analysis of the decision-making processes of the system during its operation. \item \textbf{Ethical Transparency:} The disclosure of the ethical principles and considerations that are incorporated into the system's design and operation. 
\end{enumerate}

These criteria collectively provide a framework for evaluating and integrating transparency into AI and robotic systems, encompassing both technological and ethical aspects.

The importance of transparency in ethical AI is manifold:

\begin{enumerate} \item \textbf{Enhancing Accountability:} Transparency enables the scrutiny of a system’s actions and decisions. In situations involving significant risks—such as when robotic systems make decisions that could impact human health or safety, as in medical diagnostic tools or autonomous vehicles—the ability to trace these decisions becomes critical for assigning accountability and resolving issues \cite{dignum2019responsible, awad2018moral}.
\item \textbf{Ensuring Informed Consent:} As robotic systems become more pervasive, people interacting with these systems have the right to understand how decisions affecting them are being made. This transparency is vital for maintaining human autonomy and privacy in interactions with AI \cite{jobin2019global, tegmark2017life}. However, this right is not universally applicable and needs well-defined boundaries, as outlined in Section V.

\item \textbf{Assisting Algorithmic Debugging:} Transparency of AI system decision-making helps developers to effectively identify and address biases, errors, or unintended behaviors. Such improvements are fundamental for building reliable and ethically aligned AI systems \cite{gunning2019darpa, russell2019human}.

\item \textbf{Building Public Trust:} Greater transparency can lead to increased understanding and acceptance of robotic systems, thus accelerating their adoption in various sectors \cite{hoff2015trust}.
\end{enumerate}

Despite the importance of transparency, many current robotic systems fall short in this regard. A significant number of AI algorithms, particularly those based on deep learning, are treated as ``black boxes," which means their decision-making processes remain opaque \cite{samek2019explainable}. This lack of transparency leads to numerous challenges, such as difficulties in detecting and correcting biases, challenges in verifying compliance with ethical and legal standards, reduced public trust, and obstacles in conducting effective ethical reviews and audits \cite{mehrabi2021survey, fjeld2020principled, raji2020closing}.

\section{Challenges in Implementing Transparency}
While the case for transparency in ethical AI decision-making is compelling, implementing it in practice presents several significant challenges. These can be broadly categorized into technical, ethical, and practical challenges, as summarized in Table \ref{tab:challenges}.

\begin{table}[h]
\caption{Challenges in Implementing Transparency in AI Systems}
\label{tab:challenges}
\centering
\begin{tabular}{|p{2cm}|p{2.5cm}|p{2.5cm}|}
\hline
\textbf{Category} & \textbf{Challenge} & \textbf{Implications} \\
\hline
Technical & Complexity of AI algorithms & Difficulty in providing simple explanations \\
\hline
Ethical & Privacy concerns & Balancing transparency with data protection \\
\hline
Practical & User comprehension & Conveying complex information to diverse users \\
\hline
\end{tabular}
\end{table}

\subsection{Technical Challenges}
Modern AI systems, particularly those based on transformer models and deep learning, often involve intricate architectures with billions of parameters, making it difficult to provide simple, human-understandable explanations of their decision-making processes \cite{lecun2015deep, gilpin2018explaining}. There's often a perceived trade-off between the predictive power of an AI model and its interpretability; highly accurate models tend to be more complex and less transparent, while more interpretable models may sacrifice some degree of performance \cite{adadi2018peeking, gunning2019xai}.

However, Rudin and Radin \cite{rudin2019stop} have challenged this perceived trade-off between transparency and model performance. The authors argue that interpretable models can achieve accuracy comparable to ``black box" models, questioning the necessity of using opaque AI in high-stakes domains. This perspective highlights the importance of critically examining our assumptions about the relationship between model complexity and performance.

\subsection{Challenges Specific to Robotic Systems} 
Robotic systems face unique challenges in ensuring transparency, predominantly as a result of their physical presence and their direct interactions with real-world environments:

\begin{enumerate} \item \textbf{Real-time Decision Making:} Robotic systems often operate in highly dynamic environments, requiring continuous learning and adaptation. This constant evolution adds layers of complexity in achieving transparency for these systems \cite{deng2018artificial}.In robotics, the capacity for real-time decision-making and adaptation to evolving conditions is essential. Crafting clear, understandable explanations for actions taken in such unpredictable environments requires innovative strategies that can keep pace with the rapid changes the system undergoes.
\item \textbf{Multi-modal Interactions:} Robots integrate multiple sensors and actuators, leading to decision-making processes that are inherently complex and multi-modal. These multi-layered interactions are much harder to explain than those of purely data-driven AI systems.

\item \textbf{Safety-Critical Operations:} Many robotic systems operate in areas where safety cannot be compromised. In these cases, transparency must be delicately balanced with the need for rapid, dependable performance, as revealing too much may jeopardize system responsiveness or reliability.

\item \textbf{Human-Robot Interaction:} Effectively conveying a robot's goals and decision-making processes in real-time, in a manner comprehensible to humans, continues to pose a significant challenge. This necessitates the creation of intuitive, user-friendly communication mechanisms that effectively communicate the robot's activities and intents.

\end{enumerate}

\subsection{Ethical Challenges} Although transparency is essential, it must be weighed against the need to safeguard sensitive information and protect individual privacy. In certain situations, full transparency might expose proprietary data or compromise user confidentiality \cite{wachter2017counterfactual, lepri2018fair}. Providing detailed explanations of a system's decision-making process could also be misused by malicious actors to manipulate or exploit the system \cite{amodei2016concrete}.

Additionally, efforts to make complex systems easier to understand carry the risk of oversimplification, which may result in misunderstandings or misplaced confidence \cite{miller2019explanation}. Finding the right balance between providing enough detail and ensuring comprehensibility remains a significant challenge, especially when addressing the varied needs of different stakeholders, ranging from technical experts to end-users.

\subsection{Practical Challenges} Achieving transparency on a technical level is only part of the challenge; effectively communicating this information in a way that is clear and meaningful to users with diverse levels of technical expertise remains difficult \cite{hoffman2018metrics}. In dynamic settings where robots need to make split-second decisions, providing real-time explanations without compromising system performance is particularly challenging \cite{anjomshoae2019explainable}.

Implementing transparency initiatives may also demand extra computational resources, which can affect the cost-effectiveness and efficiency of robotic systems \cite{ribeiro2016trust}. Moreover, promoting transparency in ethical AI involves knowledge from multiple disciplines, including computer science, ethics, cognitive science, and human-computer interaction, making it a complex, interdisciplinary issue \cite{doshi2017accountability}.

\section{Proposed Approaches to Enhance Transparency}
To address the challenges outlined in the previous section and move towards more transparent ethical AI decision-making in robotic systems, we propose several approaches. Table \ref{tab:approaches} provides a comparison of these transparency approaches, highlighting their advantages and challenges.

\begin{table}[h]
\caption{Comparison of Transparency Approaches}
\label{tab:approaches}
\centering
\begin{tabular}{|p{2cm}|p{2.5cm}|p{2.5cm}|}
\hline
\textbf{Approach} & \textbf{Advantages} & \textbf{Challenges} \\
\hline
Standardized Metrics & Quantifiable and comparable across systems & Difficult to standardize across diverse AI applications \\
\hline
XAI Techniques & Provides insights into complex models & May reduce model performance \\
\hline
User-Friendly Interfaces & Improves user understanding and trust & Requires significant design effort \\
\hline
Transparency-by-Design & Proactive approach, Integrates ethics early & May slow initial development process \\
\hline
\end{tabular}
\end{table}

\subsection{Targeted Transparency Approaches for Different Stakeholders}
The proposed transparency measures cater to both system designers and end-users. For system designers, the focus is on providing in-depth metrics and insights for debugging and improvement. This includes detailed information about the AI models, training data, and decision-making processes. For end-users, emphasis is placed on providing intuitive, user-friendly explanations that convey the system's capabilities, limitations, and basic decision-making rationale without requiring technical expertise.

\subsection{Developing Standardized Transparency Metrics}
Creating a comprehensive transparency index that quantifies the level of transparency in a robotic system, considering factors such as explainability, interpretability, and accessibility of information, is crucial \cite{mohseni2021multidisciplinary}. This should be complemented by collaborating with standards organizations to create widely accepted transparency benchmarks for different types of robotic systems and applications \cite{brundage2020toward}. Implementing a framework for periodic assessments of robotic systems against these standardized metrics can ensure ongoing compliance and improvement \cite{raji2019actionable}.

\subsection{Incorporating Explainable AI (XAI) Techniques}
Where possible, AI models that are intrinsically more interpretable, such as decision trees or rule-based systems, should be used for critical ethical decision-making components \cite{rudin2019black}. For complex models like deep neural networks, techniques such as Local Interpretable Model-agnostic Explanations (LIME) or SHapley Additive exPlanations (SHAP) can give insights into decision-making processes \cite{lundberg2017unified}. Developing hybrid systems that combine the power of complex AI models with more transparent, rule-based systems for ethical decision-making is another promising approach \cite{dosilovic2018explainable}.

\subsection{Creating User-Friendly Interfaces}
Designing interfaces that provide explanations through various modalities (e.g., visual, textual, auditory) can cater to different user preferences and cognitive styles \cite{abdul2018trends}. Implementing AI-driven interfaces that adjust the level and complexity of explanations based on the user's expertise and context can enhance understanding \cite{wang2019designing}. Developing tools that enable users to explore the decision-making process interactively, allowing them to ask questions and receive relevant explanations, can further improve transparency \cite{weld2019challenge}.

\subsection{Establishing Transparency Requirements in Design}
Integrating transparency considerations from the earliest stages of robotic system design, making it a fundamental requirement rather than an afterthought, is essential \cite{dignum2018ethics}. Conducting thorough ethical impact assessments during the design phase can help identify potential ethical issues and transparency needs \cite{wright2013integrating}. Involving diverse stakeholders, including ethicists, end-users, and policymakers in the design process can ensure transparency measures meet varied needs and expectations \cite{hagendorff2020ethics}.

\section{Implications and Future Directions}
Prioritizing transparency in ethical AI decision-making for robotic systems has far-reaching implications and opens up several avenues for future research and development. Table \ref{tab:future_research} outlines key research areas and associated questions for future work in AI transparency.

\begin{table}[h]
\caption{Future Research Directions in AI Transparency}
\label{tab:future_research}
\centering
\begin{tabular}{|p{2.5cm}|p{5cm}|}
\hline
\textbf{Research Area} & \textbf{Key Questions and Objectives} \\
\hline
Cognitive Models of Explanation & How do humans process and understand AI explanations? \\
\hline
Multi-Agent Transparency & How can transparency be maintained in systems with multiple AI agents? \\
\hline
Long-term Impact Studies & What are the long-term effects of increased AI transparency on public trust? \\
\hline
AI Literacy Programs & How can we effectively educate the public about AI decision-making? \\
\hline
Domain-Specific Transparency & What are the unique transparency needs in healthcare, autonomous vehicles, etc.? \\
\hline
\end{tabular}
\end{table}

\subsection{Impact on Public Trust and Adoption}
Increased transparency can lead to greater understanding and acceptance of robotic systems, potentially accelerating their adoption in various sectors \cite{de2018automation}. Transparent systems may facilitate more effective teamwork between humans and robots, as humans can better understand and predict robot behavior \cite{shneiderman2020human}. Moreover, transparency metrics could enable consumers to make more informed decisions about the robotic products and services they use, driving market demand for ethical AI \cite{floridi2019establishing}.

\subsection{Implications for Regulatory Frameworks}
Transparent AI systems provide policymakers with clearer insights into robot decision-making, enabling more informed and effective regulation \cite{cath2018artificial}. The push for transparency could drive international efforts to standardize ethical AI practices, similar to existing standards in other technological domains \cite{theodorou2020towards}. Enhanced transparency may contribute to creating more sophisticated legal frameworks for determining liability in situations involving autonomous robotic systems\cite{kingston2018artificial}.

\subsection{Ethical Reflections on the Right to Explanation} 
The ethical duty to provide transparency must take into account the right to explanation, similar to human decision-making processes. However, it is essential to critically evaluate the extent and limits of this right in interactions between humans and AI. Unlike interactions between humans, where such rights may vary by context, interactions between humans and robots often require a higher level of clarity, especially in safety-critical situations.

The right to explanation is particularly vital in fields such as healthcare or criminal justice, where AI decisions significantly impact individuals' lives. Nevertheless, this right may not always be applicable or absolute. Key factors that influence the scope of this right include:

\begin{itemize} 
\item The potential effect of the decision on human lives and well-being 
\item The complexity involved in the decision-making process 
\item The urgency of the decision 
\item Considerations related to privacy and security 
\item The technical feasibility of delivering an understandable explanation 
\end{itemize}

Continued research and ethical discussions are necessary to establish clear guidelines regarding when and how this right should be implemented in different human-AI interaction scenarios.

\subsection{Future Research Directions}
Several key areas for future research emerge from our analysis:

\begin{enumerate}
    \item \textbf{Cognitive Models of Explanation:} Further research into how humans process and understand explanations can inform the development of more effective transparency mechanisms \cite{miller2019explanation2}.
    
    \item \textbf{Multi-Agent Transparency:} Exploring how to maintain transparency in complex scenarios involving multiple interacting robotic agents is another crucial area of study \cite{dafoe2020open}.
    
    \item \textbf{Long-term Impact Studies:} Longitudinal studies on how increased transparency affects public perception, trust, and interaction with robotic systems over time are needed \cite{cave2019scary}.
    
    \item \textbf{AI Literacy Programs:} Developing educational programs to improve public understanding of AI decision-making processes and ethical considerations is essential \cite{long2020ai}.
    
    \item \textbf{Domain-Specific Transparency:} Investigating transparency requirements and solutions for specific applications of robotic systems, particularly in areas like healthcare, autonomous vehicles, and smart cities, will be crucial for the ethical development of these technologies \cite{trocin2023responsible}.
\end{enumerate}

\section{Conclusion}
This paper has advocated for transparency as a fundamental ethical principle in AI decision-making for robotic systems, underpinning trust, accountability, and responsible innovation. We've proposed a comprehensive framework addressing technical and ethical aspects, including standardized metrics, explainable AI techniques, user-friendly interfaces, and design-phase transparency requirements. The complex relationship between transparency and trust necessitates coupling transparency efforts with broader ethical considerations and stakeholder dialogue.

We urge the robotics and AI community to prioritize transparency through interdisciplinary collaborations. Future work should focus on developing standardized metrics, advancing explainable AI, improving human-robot interaction interfaces, and researching long-term impacts on public trust and AI adoption. By elevating transparency to a core principle, we can encourage responsible technological advancement, realizing AI and robotics' potential to improve human life while ensuring accountability and alignment with human values.

\bibliographystyle{IEEEtran}
\bibliography{root}

\end{document}